\documentclass[final,times,5p,twocolumn]{elsarticle}

\usepackage{amssymb}
\usepackage{lipsum}
\usepackage{graphicx}
\usepackage{dcolumn}
\usepackage{bm}
\usepackage{soul}
\usepackage{slashed}
\usepackage{physics}
\usepackage{amsmath}
\usepackage{placeins}
\usepackage[normalem]{ulem}
\usepackage{mathtools,slashed}
\usepackage{hyperref}
\biboptions{sort&compress} 
\usepackage{xcolor}
\usepackage[table]{xcolor}

\journal{Physics Letters B}

\begin{document}

\begin{frontmatter}



\title{Hidden Supersymmetry in  AdS/QCD models for Fermions}


\author{Rafael A. Costa-Silva$^*$}
\ead{rc-fis@outlook.com, rafaelcosta@pos.if.ufrj.br}
\cortext[mycorrespondingauthor]{Corresponding author}
\author{Henrique Boschi-Filho}
\ead{boschi@if.ufrj.br}
\address{Instituto de Fisica, Universidade Federal do Rio de Janeiro, Rio de Janeiro,
            21.941-909, Brazil}

\begin{abstract}
 We show that the chiral modes splitted from the Dirac equation satisfying Schrodinger-like equations in the context of deformed AdS/QCD model are isospectral, since the chiral effective potentials  are SUSY partners. In particular, we show how to perform truncations of these potentials that preserve isospectrality of the chiral modes. Following this prescription, we compare a quadratic approximation of the deformed AdS model with the dressed soft wall model for fermions and show how the spectra of both models can coincide.
\end{abstract}




\end{frontmatter}




\section{Introduction}
\label{introduction}

The description of strong interactions is usually made by QCD, which works very well in high-energy set-ups. Despite that, for low-energy systems, QCD becomes non-perturbative and other resources and techniques must be implemented. An AdS/CFT-based approach is the so-called AdS/QCD, a bottom-up technique  which describes some systems, governed by strong interactions, using known properties of QCD as a guideline.

    Some AdS/QCD models have been developed, in particular we may cite the soft wall \cite{Karch:2006pv, Kim:2007rt, Colangelo:2007pt, Grigoryan:2007my}, the light-front holographic QCD \cite{Brodsky:2006uqa, Brodsky:2014yha,  deTeramond:2014asa}, and the deformed AdS model \cite{Andreev:2006vy, Andreev:2006ct,  FolcoCapossoli:2019imm, Nascimento:2023dzx}. See also \cite{Vega:2008af, Colangelo:2008us, Abidin:2009hr, Herzog:2006ra, Csaki:2006ji, Nakano:2006js,  Kim:2006uta, BallonBayona:2007vp,  Kwee:2007dd, He:2007juu, BallonBayona:2007qr,  Forkel:2007ru,  Batell:2008zm, Dubynskiy:2008di, dePaula:2008fp, Dietrich:2008up,  Forkel:2008un, Gursoy:2009zza,   Miranda:2009uw, Fujita:2009ca,  Afonin:2010fr, Branz:2010ub, Karch:2010eg, Gutsche:2011vb, He:2013qq, Li:2013oda,   Dudal:2015wfn, Rodrigues:2016kez, Mamo:2019mka} for further developments of the soft wall model and \cite{Andreev:2006eh, Andreev:2006nw, Andreev:2007vn, Kajantie:2006hv, White:2007tu, Csaki:2008dt, Giannuzzi:2009gh, Andreev:2009zk, Colangelo:2010pe, Li:2011hp, Yang:2015aia, Bernardini:2016hvx, Arefeva:2018hyo, Bruni:2018dqm, Giannuzzi:2019esi,  Bohra:2019ebj,  
    FolcoCapossoli:2020pks, Contreras:2021epz, MartinContreras:2021yfz,  Nascimento:2025smf, Issifu:2026nay, Wang:2026dcy, Ovchinnikov:2026mve} for interesting applications and/or extensions of the deformed AdS model. 
    
    The soft wall model consists of the introduction of a quadratic dilaton in the five-dimensional bulk action presenting analytical solutions and leads to linear Regge trajectories for mesons \cite{Karch:2006pv}. In order to construct a quark-antiquark potential the quadratic dilaton was alternatively introduced as a warp factor in the original AdS metric leading to the now called deformed AdS model which only has analytical solutions for mesons \cite{Andreev:2006vy, Andreev:2006ct}.
    
    The extension of the soft wall for baryons was presented in \cite{Abidin:2009hr} and requires the introduction of a potential in the mass term in the Dirac action in order to bring a discrete spectrum. This model is usually called as the dressed soft wall model, since the mass is corrected by an energy (position) dependent potential, in analogy with a renormalized mass. In the case of the deformed AdS model for baryons the original model need not to be modified but it requires numerical solutions \cite{FolcoCapossoli:2019imm}. Baryon spectra was also discussed in light-front holographic QCD  \cite{Brodsky:2006uqa, Brodsky:2014yha} and its connection with supersymmetric quantum mechanics (SUSYQM) \cite{Witten:1981nf, Cooper:1994eh} 
    was proposed in \cite{deTeramond:2014asa}. 
    
    In order to obtain the spectrum of  fermions holographically, one usually writes the Dirac action in a convenient higher dimensional model from which is deduced the corresponding Dirac equation for spinors which are then splitted in chiral left- and right-modes. This leads to coupled equations of motion that can be decoupled to obtain Schrödinger-like equations with corresponding chiral potentials (see, {\sl e.g.} \cite{Abidin:2009hr,FolcoCapossoli:2019imm, Nascimento:2023dzx} for this procedure in the soft wall and deformed AdS models). 

    A necessary consistency condition is that both chiral modes present the same mass spectrum as they are just a split of the same physical object. In this work, we show that this is so in the deformed AdS model because in general the obtained chiral potentials that appear in the Shrödinger-like equations are SUSY partners and therefore, isospectral. As the solutions for the chiral modes in the deformed AdS model require numerical analysis, it is sometimes convenient to resource to approximations of the corresponding potentials. However, it is in general not obvious how to implement these approximations in order to preserve the isospectrality of the chiral modes. Here, we develop a systematic way to obtain a consistent truncation of the deformed AdS potentials that keeps the chiral modes isospectral. We also show that it is possible to relate the spectra of the dressed soft wall model for fermions with the ones from a quadratic approximation of the deformed AdS model, establishing an equivalence of these models in this regime. 

The remainder of this paper is organized as follows. In Section \ref{DAdSmodel}, we present the deformed AdS model for fermions and obtain the chiral decomposition satisfying Schrödinger-like equations with the corresponding chiral potentials. In Section \ref{SUSY}, we define the superpotential and show its relation with the chiral potentials then implying their isospectrality. In Section \ref{Trunc}, we give a prescription to consistently truncate the fermionic potentials of the deformed AdS model to ensure the preservation of the isospectrality of the chiral modes and consider the case of the truncation of the superpotential with order ${\cal O}(z^3)$ implying chiral potentials with order ${\cal O}(z^6)$ which we solve numerically. In Section \ref{Comp}, we compare the fermionic spectra obtained from the dressed soft wall and from the quadratic approximation of the deformed AdS model  following the prescription for isospectrality found in the previous section. Then, in Section \ref{Conclusions}, we present our final comments and conclusions.


\section{Deformed AdS model}
\label{DAdSmodel}

 We start by defining a deformed AdS$_5$ space, with radius $R=1$, which metric is given by \cite{FolcoCapossoli:2020pks}
    \begin{equation}
        ds^2=g_{mn}dx^{m}dx^{n}=e^{2A(z)}(dz^2+\eta_{\mu\nu}dy^{\mu}dy^{\nu}).
    \end{equation}
    Here, $m,$ $n$ refers to indices in the 5-dimensional space; $\mu,$ $\nu$ are indices of the boundary four dimensional Minkowski space with metric $\eta_{\mu\nu}=\rm{diag}(-,+,+,+)$, and $z$ is the so-called holographic fifth coordinate. The warp factor $A(z)$ in the metric gives us the deformed AdS$_5$
    \begin{equation}\label{warp}
        A(z)=-\log{z}+\frac{1}{2}kz^2.
    \end{equation}
    The constant $k$ has mass dimension 2 and is associated with some QCD mass scale and one can recover the pure AdS space in the limit of $k\rightarrow0$. In order to describe fermions, we start with an action for fermionic fields in the deformed AdS$_5$ space
    \begin{equation}
        S=\int{d^5x\sqrt{-g}\,\bar{\Psi}(\slashed{D}-m_5)\Psi},
    \end{equation}
    with covariant derivative contracted with the Dirac matrices given by \cite{FolcoCapossoli:2020pks}
    \begin{eqnarray}
\slashed{D}&=&g^{mn}e^{a}_{n}\gamma_{a}\bigg(\partial_{m}+\frac{1}{2}\omega^{bc}_{m}\Sigma_{bc}\bigg) \cr 
&=& e^{-A(z)}\gamma^{5}\partial_{5}+e^{-A(z)}\gamma^{\mu}\partial_{\mu}+2A'(z)\gamma^{5}.
    \end{eqnarray}

    From this action, one can obtain the equation of motion, that can be written as
    \begin{equation}
        [e^{-A(z)}\gamma^{5}\partial_{5}+e^{-A(z)}\gamma^{\mu}\partial_{\mu}+2A'(z)\gamma^{5}-m_5]\Psi=0,
    \end{equation}
    where $\partial_5 = \partial_z$, and $m_5$ is the baryon bulk mass. Decomposing the spinor $\Psi$ into right- and left-handed chiral components, one has
    \begin{equation}
        \Psi(x^{\mu},z)=\bigg[\frac{1-\gamma^{5}}{2}f_{L}(z)+\frac{1+\gamma^{5}}{2}f_{R}(z)\bigg]\Psi_{(4)}(x), \label{Psi}
    \end{equation}
    where $\Psi_{(4)}(x)$ satisfies the usual Dirac equation $(\slashed{\partial}-M)\Psi_{(4)}(x)=0$ in flat four-dimensional spacetime. Notice that $M$ is the four-dimensional fermionic mass, and the chiral modes satisfy the relations
    \begin{equation}
        \gamma^{5}f_{L/R}=\mp f_{L/R},
        \label{relf1}
    \end{equation}
    and
    \begin{equation}
        \gamma^{\mu}\partial_{\mu}f_{R}=M f_{L}.
        \label{relf2}
    \end{equation}

    A natural consequence of this construction is that the chiral modes have the same mass, as a consistency condition. Considering that the Kaluza-Klein modes are dual to chiral spinors, one can expand
    \begin{equation}
        \Psi_{L/R}(x^{\mu},z)=\sum_{n}f^{n}_{L/R}(x^{\mu})\, \chi^{n}_{L/R}(z).
    \end{equation}

    So, we can obtain the coupled equations
    \begin{equation}
        (\partial_{z}+2A'(z)e^{A(z)}+m_{5}e^{A(z)})\chi^{n}_{L}(z)=M_{n}\, \chi^{n}_{R}(z)
    \end{equation}
    and
    \begin{equation}
        (\partial_{z}+2A'(z)e^{A(z)}-m_{5}e^{A(z)})\chi^{n}_{R}(z)=-M_{n}\, \chi^{n}_{L}(z),
    \end{equation}
    where $M_n$ are the four-dimensional fermionic masses of both left and right $\chi^{n}_{L/R}(z)$ modes. 

    Performing the Bogoliubov transformations $\chi^{n}_{L/R}(z)=\psi^{n}_{L/R}(z) \, e^{-2A(z)}$, and decoupling the equations for the chiral modes, we obtain the  Schrödinger-like equations
    \begin{equation}\label{schrodingereq}
        \bigg[-\frac{d^2}{dz^2}+V_{R/L}(z)\bigg]\, \psi^{n}_{R/L}(z)=M_{n}^2\,\psi^{n}_{R/L}(z),
    \end{equation}
    where the effective chiral potentials are given by
    \begin{equation}\label{potentials}
        V_{R/L}(z)=m_5^2e^{2A(z)}\pm m_5e^{A(z)}A'(z).
    \end{equation}

    It is natural to believe that both modes have the same spectrum, since this is a consistency condition, implied by Eqs. \eqref{Psi}-\eqref{relf2}. However, it is not obvious from the potentials in (\ref{potentials}) that they produce the same spectrum in equation (\ref{schrodingereq}). 
    
    In next section, we show how to relate this problem to SUSY-QM and prove the isospectrality of the chiral modes. We also show how to perform consistent truncations in these potentials while keeping the isospectrality of the chiral modes.


\section{SUSY-like construction}
\label{SUSY}

    We start noting that we can define the so-called superpotential \cite{Cooper:1994eh, Witten:1981nf}
    \begin{equation}\label{superpotential}
        W(z)=m_{5}e^{A(z)},
    \end{equation}
    with derivative 
    \begin{equation}
        W'(z)=m_5e^{A(z)}A'(z).
    \end{equation}

    With these definitions, the chiral potentials in equation (\ref{potentials}) may be written as
    \begin{equation}\label{potentialrel}
        V_{R/L}(z) = W^2(z)\pm W'(z).
    \end{equation}

    This relation, as we will see, will imply the condition that the left and right potentials must satisfy to be called supersymmetric partners. 
    
    Let us also define the operator
    \begin{equation}\label{Q}
        Q=\frac{d}{dz}+W
    \end{equation}
    and its adjoint
    \begin{equation}\label{Qdagger}
        Q^{\dagger}=-\frac{d}{dz}+W.
    \end{equation}

    With these definitions, we can see that the left and right Hamiltonians $H_{R/L}\equiv(-d^2/dz^2 + V_{R/L}$) that appear in equation (\ref{schrodingereq}) may be written as
    \begin{equation}
        H_{R}=Q Q^{\dagger}
    \end{equation}
and
\begin{equation}
    H_{L}=Q^{\dagger}Q.
\end{equation}

    Since $H_{R}^{\dagger}=H_{L}$, we can see that they are isospectral. Let us prove this assuming that there is no zero energy mode. 
    
    So, supose that the left and right ground state wave functions $\psi^{0}_{L}$ and $\psi^{0}_{R}$ energies respectively satisfy $E^{0}_{L}>0$ and $E^{0}_{R}>0$. Then we have
    \begin{equation}\label{eq1}
        H_{L}\psi^{0}_{L}=E^{0}_{L}\psi^{0}_{L},
    \end{equation}
    and 
    \begin{equation}
        H_{R}\psi^{0}_{R}=E^{0}_{R}\psi^{0}_{R}.
    \end{equation}
    From equations (\ref{Q}) and (\ref{Qdagger}), we can see that
    \begin{equation}
        H_{R}Q=QH_{L},
    \end{equation}
    which leads us to
    \begin{equation}\label{eq2}
        H_{R}(Q\psi^{0}_{L})=Q(H_{L}\psi^{0}_{L})=E^{0}_{L}(Q\psi^{0}_{L}).
    \end{equation}
    Since $E^{0}_{L}>0$, we have
    \begin{equation}
        ||Q\psi^{0}_{L}||^2=\bra{\psi^{0}_{L}Q^{\dagger}}\ket{Q\psi^{0}_{L}}=\bra{\psi^{0}_{L}}\ket{H_{L}\psi^{0}_{L}}=E^{0}_{L}||\psi^{0}_{L}||^2>0.
    \end{equation}
    This relation ensures that
    \begin{equation}
        Q\psi^{0}_{L}\neq 0.
    \end{equation}

    So, we can see that $E^{0}_{L}$ is eigenvalue of $H_{R}$, which implies the relation
    \begin{equation}\label{rel1}
    E^{0}_{L}\geq E^{0}_{R}. 
    \end{equation}

    The same procedure may be used to show that
    \begin{equation}\label{rel2}
        E^{0}_{R}\geq E^{0}_{L}.
    \end{equation}

    So, comparing equations (\ref{rel1}) and (\ref{rel2}), if there is no zero energy mode, i.e., if the left and right ground states are not massless (a reasonable physical coinstraint, since our fermions in the boundary are not massless), the ground states of the right and left modes are equal, as expected: 
    \begin{equation}
        E_{R}^0 = E_{L}^{0}. 
    \end{equation}
    
    Nottice that this reasoning can also be applied for any excited states. Since $E_{0}^{R}=E_{0}^{L}$, we can proceed from equation (\ref{eq1}) for $\psi_{L}^1$ and $E_{L}^1$ to conclude that $E_{L}^{1}=E_{R}^{1}$, and so on. In general, if we apply this method, we can prove the isospectrality for any excitation $n$. This is so, because  equation (\ref{eq2}) also holds for any excited states as, 
    \begin{equation}
        H_{R}(Q\psi_{L}^{n})=E_{L}^{n}(Q\psi_{L}^{n}).
    \end{equation}
      
    In this way, $(Q\psi_{L}^{n})$ is an eigenstate of $H_{R}$ with eigenvalue $E_{L}^{n}$. If the eigenvalues are ordered as $E_{L}^{0}<E_{L}^{1}<...E_{L}^{n}...$, and the same for right eigenvalues, so we can compare any excitation. As this condition is true, they are isospectral for any excitation $n$.
    \begin{equation}
        E_{R}^{n}=E_{L}^{n}=E^{n}
    \end{equation}
    Note that SUSY partners potentials implies in isospectrality for these  modes. 
    
    So, we will search for potentials that satisfies this SUSY like relation, so we ensure isospectrality. In general, if we perform expansions directly in the potentials given by (\ref{potentials}), the SUSY partner relation disappears. Alternatively, we will show how to perform approximations in these potentials that preserve the isospectrality of chiral modes in the deformed AdS model in the following section.


    \section{A systematic truncation that preserves isospectrality}
    \label{Trunc}

Since in the example discussed so far of the deformed AdS model, it requires numerical solutions for the chiral modes, it is natural to ask what approximations for the potentials are useful in possible applications. Then, in the light of the results of the previous section, we wonder what truncations of the potentials (\ref{potentials}) could preserve the isospectrality of the approximate chiral modes. 

In general, these truncated potentials will not be SUSY partners and consequently not isospectral. So, if we want isospectrality between  chiral modes described by approximate potentials, we need to ensure that the SUSY partner relation holds for some superpotential. Then,  we will expand the original superpotential $W(z)$ up to some order in $z$ and obtain the corresponding approximate chiral potentials  that, by construction, will be SUSY partners and therefore, isospectral. 

 Using the definition of the superpotential (\ref{superpotential}) in our example of the deformed AdS model, with $A(z)$ defined in equation (\ref{warp}), we see that $W(z)$ only presents odd powers of $z$, 
  \begin{equation}\label{W(z)}
        W(z)=\frac{m_5}{z}+\frac{m_5k}{2}z+\frac{m_5k^2}{8}z^3 + \frac{m_5k^3}{48}z^5 + \cdots
    \end{equation}
 
 So, truncating this expansion up to $\mathcal{O}(z)$, one obtains 
    \begin{equation}
        W^{[1]}(z)=\frac{m_5}{z}+\frac{m_5k}{2}z.
    \end{equation}
    From this expression, we see that the corresponding chiral potentials are
    \begin{equation}\label{goodpotential1}
        V^{[1]}_{R/L}(z)=\frac{m_5(m_5\mp1)}{z^2}+km_{5}\bigg(m_5\pm\frac{1}{2}\bigg)+\frac{m_5^2k^2}{4}z^2.
    \end{equation}

    By construction, since these potentials satisfy the relation (\ref{potentialrel}), they are SUSY partners, and therefore, isospectral. In fact, the Schrödinger equation for these potentials has analytical solutions and the spectra are given by
    \begin{equation}\label{deformedspectra}
        M_{n}^2=2m_5k\bigg(n+m_5+\frac{1}{2}\bigg); \qquad (n=0,1,2,\cdots). 
    \end{equation}

    In a recent paper \cite{Nascimento:2023dzx}, the authors assumed that the truncation could be done directly on $V_{R/L}(z)$. However, if one takes  the function $A(z)$ in (\ref{potentials}) and expand these potentials up to $\mathcal{O}(z^2)$, one would obtain 
    \begin{equation}\label{issue}
        \tilde{V}_{R/L}=\frac{m_5(m_5\mp 1)}{z^2}+k m_5 \bigg(m_5 \pm \frac{1}{2}\bigg)+k^2m_5\bigg(\frac{m_5}{2}\pm\frac{3}{8}\bigg)z^2.
    \end{equation}

    In fact, these potentials are also analytical, and the authors show that they do not reproduce the same spectra for both chiral modes. This issue is related to the truncation: it does not preserves the relation expected in (\ref{potentialrel}). So the right and left potentials are not SUSY partners, and therefore not isospectral. 
    
    Let us also show that they are not SUSY partners. For any $V_{R}$ and $V_{L}$ that are SUSY partners, from equation (\ref{potentialrel}), we have
    \begin{equation}\label{wrel}
        W^2=\frac{V_{L}+V_{R}}{2}\quad;\quad W' = \frac{V_{R}-V_{L}}{2}.
    \end{equation}

    Using the second identity in (\ref{wrel}), one has 
    \begin{equation}
       W(z)= \int{\bigg(\frac{V_{R}-V_{L}}{2}\bigg)\,dz}+C.
    \end{equation}
    Using this result in the first identity in (\ref{wrel}) one gets 
    \begin{equation}\label{relation}
        \bigg[\int{\bigg(\frac{V_{R}-V_{L}}{2}\bigg)\, dz}+C\bigg]^2 = \frac{V_{R}+V_{L}}{2}. 
    \end{equation}

    Now, substituting the expression of the potentials (\ref{issue}), in the left-hand-side of the above identity, one finds: 
    \begin{align}
&\bigg[\int{\bigg(\frac{\tilde{V}_{R}-\tilde{V}_{L}}{2}\bigg)\,dz}+C\bigg]^2
        \cr
& =\bigg[m_5\bigg(\frac{1}{z}+\frac{kz}{2}+\frac{k^2z^3}{8}\bigg)+C\bigg]^2 
        \label{lhs}
    \end{align}
    On the other side, the substitution of the potentials (\ref{issue}), in the right-hand-side of (\ref{relation}), gives: 
    \begin{equation}
         \frac{\tilde{V}_{R}+\tilde{V}_{L}}{2}= 
         \frac{m_5^2}{z^2}+k m_5^2 + \frac{k^2m_5^2}{2} z^2.
         \label{rhs}
    \end{equation}
    Clearly, comparing the expressions \eqref{lhs} and \eqref{rhs}, one can see that there is no choice for the constant $C$ so that the potentials $\tilde{V}_{R/L}$ could satisfy (\ref{relation}). So, these potentials are not SUSY partners, and in general, not isospectral. 
    
    Note that, by construction, the potentials  (\ref{goodpotential1}) satisfy the expected relation (\ref{relation}). So, to guarantee the isospectrality of the left and right modes one just need to expand $W(z)$ at some definite order and then obtain the corresponding chiral potentials that emerge. In fact, for any order that we expand $W(z)$, the relation (\ref{potentialrel}) will be satisfied ensuring the isospectrality of the chiral modes.


    \begin{figure}[h!]
\vskip0.5cm 
\centering
\includegraphics[width=7cm]{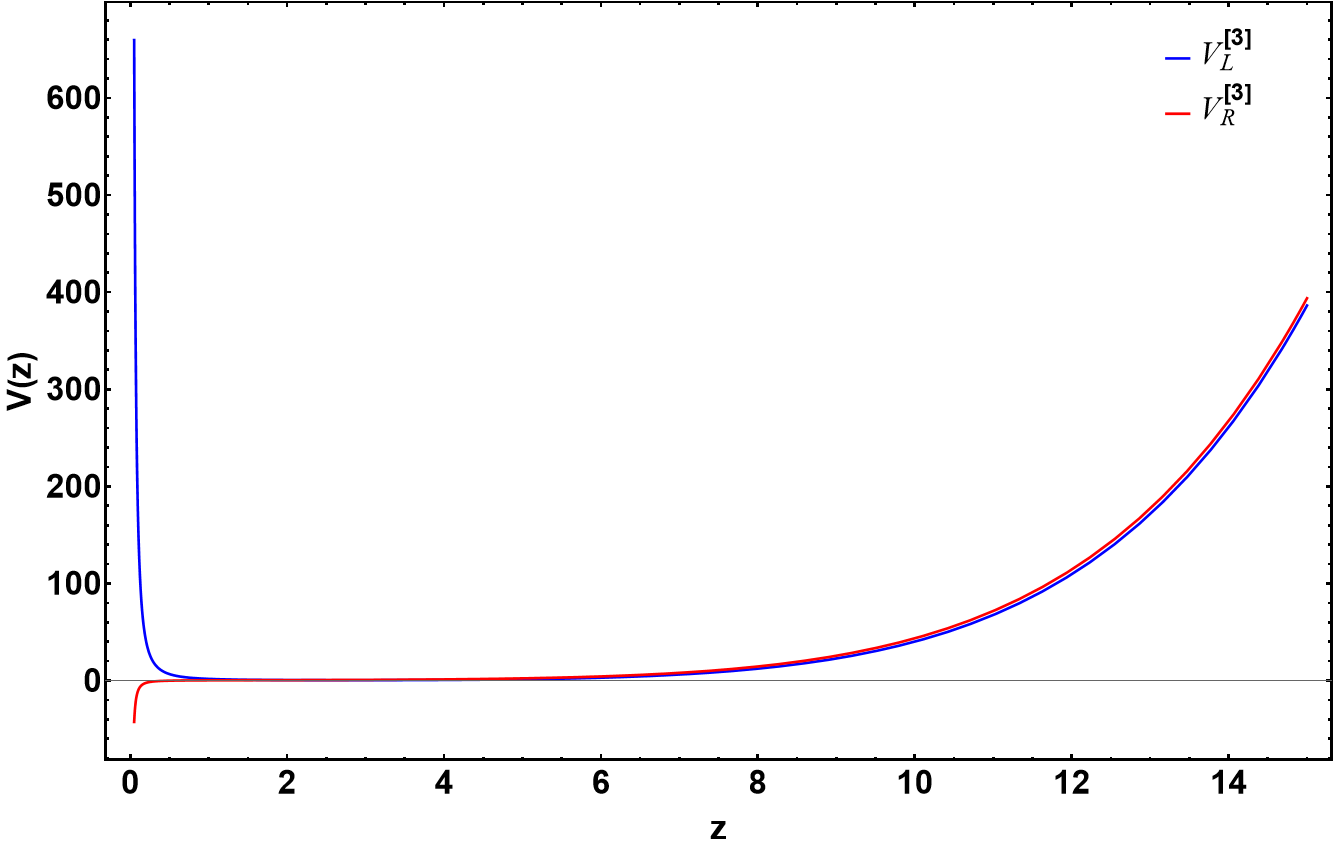}
\caption{Potentials $V^{[3]}_{R/L}(z)$ from equation (\ref{goodpotentials2}). Here, we choosed, for example, $m_5 =0.878$ GeV and $k= (0.471)^2$ GeV$^2$.}
\label{fig1potential}
\end{figure}

\begin{figure}[h!]
\vskip0.5cm 
\centering
\includegraphics[width=7.5cm]{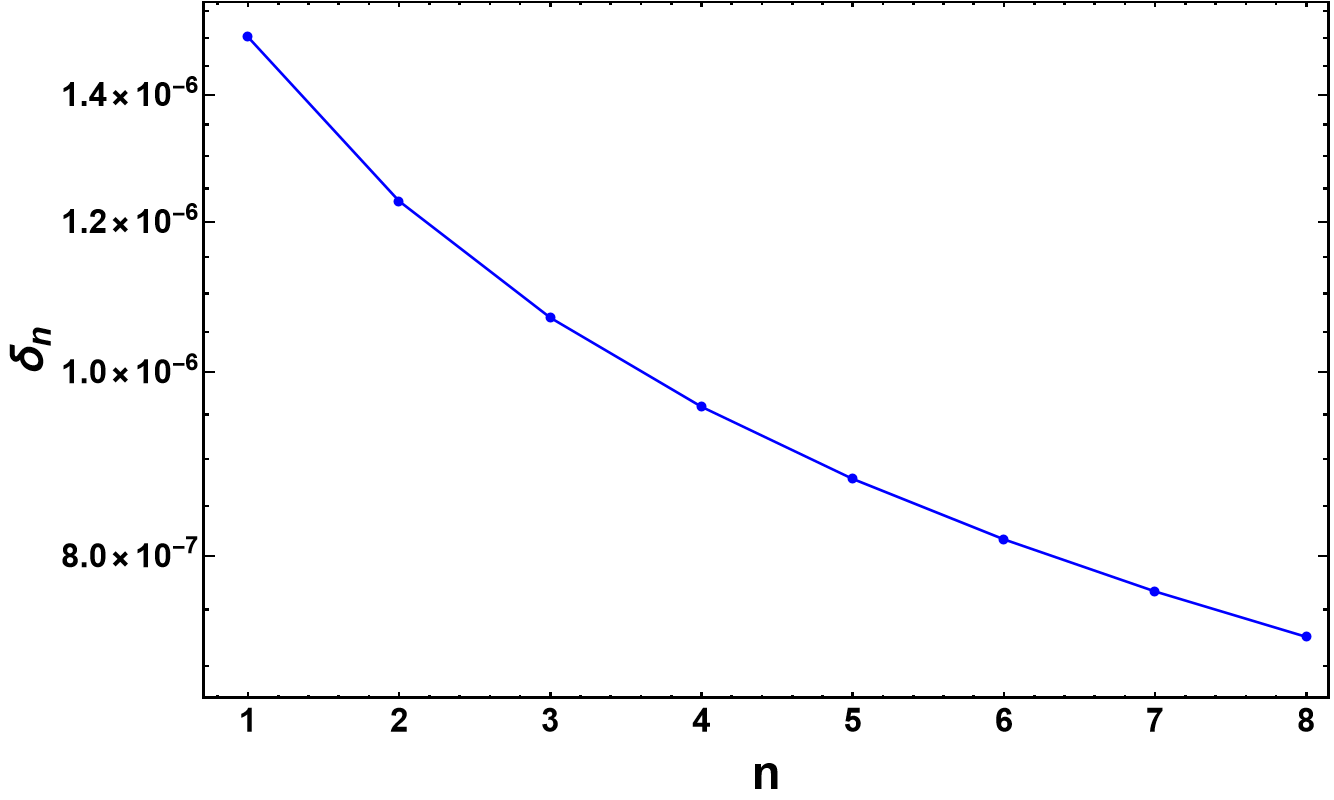}
\caption{Function $\delta_{n}$ from (\ref{error}), with masses obtained from the potentials $V^{[3]}_{R/L}(z)$ from equation (\ref{goodpotentials2}). Here, we choosed, for example, $m_5 =0.878$ GeV and $k= (0.471)^2$ GeV$^2$.}
\label{figdelta}
\end{figure}

\begin{table}[ht!]
\small
    \centering
    \begin{tabular}{|c|c|c|c|c|}
       \hline\hline
        $n$ & $M_{R}^{n}$ & $M_{L}^{n}$ & $|M_{R}^{n}-M_{L}^{n}|$ & $\delta_n$ \\
       \hline
        1 & 0.937877 & 0.937876 & $2.6 \times 10^{-6}$ & $3.0 \times 10^{-6}$ \\
        2 & 1.4138   & 1.4138   & $4.9 \times 10^{-6}$ & $2.4 \times 10^{-6}$ \\
        3 & 1.83232  & 1.83232  & $7.1 \times 10^{-6}$ & $2.1 \times 10^{-6}$ \\
        4 & 2.2169   & 2.2169   & $9.4 \times 10^{-6}$ & $1.9 \times 10^{-6}$ \\
        5 & 2.57782  & 2.57782  & $1.1 \times 10^{-5}$ & $1.7 \times 10^{-6}$ \\
        6 & 2.92082  & 2.92081  & $1.3 \times 10^{-5}$ & $1.6 \times 10^{-6}$ \\
        7 & 3.24955  & 3.24954  & $1.6 \times 10^{-5}$ & $1.5 \times 10^{-6}$ \\
        8 & 3.56651  & 3.56651  & $1.8 \times 10^{-5}$ & $1.4 \times 10^{-6}$ \\
       \hline\hline
    \end{tabular}
    \caption{Here we present the masses for the eight first chiral modes in GeV obtained from potential in (\ref{goodpotentials2}), choosing, for example, $m_5=0.878$ GeV and $k=(0.471)^2$ GeV$^2$. We also present the difference between then and the value of $\delta_{n}$ defined in equation (\ref{error}).}
    \label{tab1}
\end{table}


    Since $W(z)$ for the deformed AdS model, Eq. \eqref{W(z)}, is an odd function in $z$, it is not possible to obtain any order of $z$ in the chiral potentials $V_{R/L}$. For example, expanding $W$ up to $\mathcal{O}(z)$, we get $V_{R/L}$ of order $\mathcal{O}(z^2)$. But if we go to next order in $z$, $W$ will be like $\mathcal{O}(z^3)$, so $V_{R/L}$ will be expanded up to $\mathcal{O}(z^6)$. In fact, it is not posible with this warp factor $A(z)$, Eq. \eqref{warp}, and the SUSY like requirement, to produce, for example, $V_{R/L}$ of order $\mathcal{O}(z^4)$ or $\mathcal{O}(z^8)$. Let us see the expansion of $W(z)$ up to order $\mathcal{O}(z^3)$
    \begin{equation}
        W^{[3]}(z)=\frac{m_5}{z}+\frac{m_5k}{2}z+\frac{m_5k^2}{8}z^3,
    \end{equation}
    which leads to the approximate chiral potentials  
\begin{align}\label{goodpotentials2}
     V_{R/L}^{[3]}(z)&=\frac{m_5(m_5\mp1)}{z^2}+km_5\bigg(m_5\pm\frac{1}{2}\bigg)\cr 
         & +k^2m_5\bigg(\frac{m_5}{2}\pm\frac{3}{8}\bigg)z^2 +\frac{m_5^2k^3}{8}z^4+\frac{m_5^2k^4}{64}z^6.
\end{align}

    These potentials do not have analytical solutions, but we can see that the spectrum is equal for both modes numerically. Choosing any values for $m_5$ and $k$ we can demonstrate this for any number of modes. Let us choose $m_5=0.878$ GeV and $k=(0.471)^2$ GeV$^2$. In particular, the potentials given by equation (\ref{goodpotentials2}) are presented in Figure (\ref{fig1potential}).

    We can define some kind of relative error function to help the numerical analysis as
    \begin{equation}\label{error}
\delta_n =
\frac{\left|M_{R}^{n}-M_{L}^{n}\right|}
{\frac{1}{2}\left(M_{R}^{n}+M_{L}^{n}\right)},
    \end{equation}
 that is plotted in Log scale in Figure (\ref{figdelta}), which shows that the differences are just from numerical approximations. One can also see the masses obtained with the above potentials $V_{R/L}^{[3]}(z)$ in Table (\ref{tab1}).

    So, any order of truncation in $W(z)$ leads us to SUSY partners potentials and keeps intact the isospectrality condition.


\section{A comparison with dressed softwall}
\label{Comp}

    In the original soft wall model \cite{Karch:2006pv}, the so-called dilaton is placed in the action (not in the metric, as in deformed AdS). So, the metric is AdS
    \begin{equation}
        ds^2=g_{mn}dx^{m}dx^{n}=\frac{1}{z}(dz^2+\eta_{\mu\nu}dy^{\mu}dy^{\nu}),
    \end{equation}
    and the action for the fields become
    \begin{equation}
        S=\int{d^4x dz\sqrt{-g}e^{-\Phi(z)}\mathcal{L}},
    \end{equation}
    where $\Phi(z)=kz^2/2$ is the dilaton and $\mathcal{L}$ is the field lagrangian. Although this is a good model, for example, to describe mesonic spectra \cite{Karch:2006pv, Colangelo:2007pt, Colangelo:2008us}, this does not work for fermions. Due to the bilinear form of the Dirac action, the dilaton drops out and does not couple to fermionic fields, thus not allowing to obtain discrete spectra for fermions. An alternative model \cite{Abidin:2009hr} 
    was formulated for fermions, introducing a potential $\mathcal{U}(z)$ in the mass term, which will be interpreted as a dressed mass. Essentially, in Dirac action we make the substitution
    \begin{equation}
        m_5\rightarrow m_5+\mathcal{U}(z),
    \end{equation}
    where the potential is usually chosen to be identical with the dilaton $\Phi(z)=kz^2/2$, 
    \begin{equation}
        \mathcal{U}(z)=\frac{kz^2}{2}. 
    \end{equation} 
    Then, the Dirac action becomes
    \begin{equation}
        S=\int{d^4x dze^{-\Phi(z)}\sqrt{-g}\,\bar{\Psi}\{\slashed{D}-[m_5+\mathcal{U}(z)]\}\Psi}.
    \end{equation}
    This potential has a quadratic form in $z$ and leads to analytical solutions to the spectrum. This is sometimes called the dressed soft wall for fermions and we follow this notation here. 

    In general, for any choice of $m_5$, the chiral modes do not present the same spectrum, but for some choices of $m_5$ they are isospectral. Since we want to compare this to the quadratic approximation of deformed AdS, we will search in the region where this spectrum is real and isospectral. As the authors of \cite{Nascimento:2023dzx} showed, if $m_5\geq 1/2$, the spectrum of the chiral modes are identical and given by
    \begin{equation}\label{dressedspectra}
        M_{n}^2=2k\bigg(n+m_5+\frac{1}{2}\bigg); \qquad (n=0,1,2, \cdots). 
    \end{equation}
    So, one can conclude that this problem also has a hidden supersymmetry condition between its chiral modes. 

    We may correlate this spectrum from dressed soft wall model to the one obtained in equation (\ref{deformedspectra}) from quadratic approximation of deformed AdS. Actually, We can make both spectra equivalent. Comparing equations (\ref{dressedspectra}) and (\ref{deformedspectra}), one can see that making $m_5=1$ GeV and with an arbitrary value of $k$, the two models are isospectral for fermions. In particular, if we want to reproduce the proton mass for the ground state, we can make $k = (0.541)^2$ GeV$^2$. In this case both models are isospectral and both ground states present the mass $M_0=938$ MeV.


\section{Discussion and Conclusions}
\label{Conclusions}

    In this work we showed how the Schrödinger-like equation for the chiral modes of a spinor in the deformed AdS model naturally leads to potentials that are SUSY partners. Therefore, they are isospectral, a consistency condition, since both modes describe the same physical spinor.

    Furthermore, we showed how to implement systematical approximations in the potentials of the chiral modes from the deformed AdS model in order to ensure the SUSY partners description. This implies that one should not implement the truncation directly in the potentials $V_{R/L}(z)$, Eqs. \eqref{potentials}, at the cost of loosing the isospectrality of the chiral models. Actually, one needs to make the approximation in the so-called superpotential $W(z)$, Eq. \eqref{superpotential}, so that this will lead to approximated potentials $V_{R/L}^{[i]}$ with $i=1,3,\cdots$, Eqs. \eqref{goodpotential1} and \eqref{goodpotentials2}, that are still SUSY partners, and as a consequence, isospectral.

    Finally, we used this systematical way of truncation to obtain a quadratic approximation of $V_{R/L}(z)$ from deformed AdS space that maintains the isospectrality of the chiral modes. We compared the analytical spectrum obtained in this way to the one from the dressed soft wall, which is also a quadratic potential that presents analytical solutions. We concluded that the parameter $m_5$ can be chosen in order to guarantee the same spectrum for both models, a result not achieved before. As a future perspective of this result we may analyze the proton structure functions in DIS (deep inelastic scattering) for both models \cite{Braga:2011wa, FolcoCapossoli:2020pks} and compare them with these new tools.

\section*{Acknowledgements}
RACS is supported by Coordenação de Aperfeiçoamento de Pessoal de Nível Superior (CAPES). HBF is partially supported by Conselho Nacional de Desenvolvimento Cient\'{\i}fico e Tecnol\'{o}gico (CNPq) under grant  310346/2023-1, and Fundação Carlos Chagas Filho de Amparo à Pesquisa do Estado do Rio de Janeiro (FAPERJ) under grant E-26/204.095/2024.%


\bibliographystyle{elsarticle-num}
\bibliography{bibliography}

\end{document}